\documentclass[acmsmall]{acmart}

\AtBeginDocument{%
  \providecommand\BibTeX{{%
    \normalfont B\kern-0.5em{\scshape i\kern-0.25em b}\kern-0.8em\TeX}}}

\setcopyright{acmcopyright}
\copyrightyear{2023}
\acmYear{2023}
\acmDOI{XXXXXXX.XXXXXXX}

\acmConference[CHI '23]{April 23-28, 2023}{Hamburg, Germany}

\begin{document}

\title[Epistemic Injustices in Technology and Policy Design]{Epistemic Injustice in Technology and Policy Design: Lessons from New York City’s Heat Complaints System}

\author{Mohsin Yousufi}
\email{yousufi@gatech.edu}
\orcid{0000-0002-7738-4087}
\affiliation{%
  \institution{Georgia Institute of Technology}
  \city{Atlanta}
  \state{Georgia}
  \country{USA}
}

\author{Charlotte Alexander}
\email{calexander@gsu.edu}
\affiliation{%
  \institution{Georgia State University}
  \city{Atlanta}
  \state{Georgia}
  \country{USA}
}

\author{Nassim Parvin}
\email{nassim@gatech.edu}
\affiliation{%
  \institution{Georgia Institute of Technology}
  \city{Atlanta}
  \state{Georgia}
  \country{USA}
}


\begin{abstract}

This paper brings attention to epistemic injustice, an issue that has not received much attention in the design of technology and policy. Epistemic injustices occur when individuals are treated unfairly or harmed specifically in relation to their role as knowers or possessors of knowledge. Drawing on the case of making heat complaints in New York City, this paper illustrates how both technological and policy interventions that address epistemic injustice can fail or even exacerbate the situations for certain social groups, and individuals within them. In bringing this case to the workshop, this paper hopes to provide another generative and critical dimension that can be utilised to create better technologies and policies, especially when they deal with diverse and broad range of social groups.

\end{abstract}



\maketitle

\section{Introduction}
In January of 2022, an apartment fire in the Bronx took 17 lives, including 8 children, while dozens of others were injured \cite{miller_new_2022}. This tragedy was caused by a malfunctioning space heater. The use of ad hoc measures such as space heaters or ovens is quite common in New York City due to the large numbers of tenants who live without heat \cite{aponte_as_2022,lennard_bronx_2022}. In fact, heating complaints is the third largest number of complaints that the city registers after parking and noise complaints \cite{new_york_city_comptroller_turn_2023}. Even though New York city mandates temperature limits during the cold season that runs from October 1 to May 31, not all landlords abide by them. For most of the tenants, these frigid temperatures in their homes last for weeks, months or even years on end; the situation is endemic to all 5 boroughs in New York \cite{renthop_bronx_2022}. In the case of a non-responsive, non-complying landlord, the tenants have to reach out to the City to help them get the heat restored. This involves a complex and fraught process of navigating the City’s bureaucratic, administrative structure and the Housing Court. It is a frustrating experience for the tenants who more often than not end up without any resolution to their problem \cite{blankley_fight_2016}. There are many historical, political and economical factors that contribute to this but part of the problem stems from the inability of the tenants to “prove” the lack of heat or evidence for landlord's negligence. 

Recognizing the ubiquity and magnitude of the problem the non-profit Heat Seek, developed a technological intervention, an IoT temperature sensor that documents heat violations in a home. The sensors been highly effective at combating this lack of heat and forcing landlords to turn up the heat. This intervention, currently applied to 58 buildings \cite{heat_seek_annual_2023}, has helped tenants avoid Housing Court altogether and, according to Heat Seek's annual reports, made the landlords more cautious in cutting off heat \cite{heat_seek_after_2016}. By allowing the tenants to document the heating violations in their homes, Heat Seek has made the tenants more credible.

Why is Heat Seek so effective at holding landlords accountable? We argue that what Heat Seek does, by means of providing quantitative data, is giving tenants a "credibility boost", thus making their complaint legitimate and legible in the eyes of attorneys, landlords and the Court. This occurs because, inherently, the situation of an unresolved, unheard heating complaint stems not only from institutional mechanics \cite{ahmed_complaint_2021} but from epistemic injustice. In the following sections, the paper will briefly explore the notion of epistemic injustice as theorized by Miranda Fricker. By following the process of a heat complaint, it will then illustrate how the existing policy and legal frameworks discount the tenants knowledge. The paper will then use Heat Seek to show how a technology can respond to such epistemic injustices and also not necessarily resolve it. We close with a brief discussion on how consideration for epistemic injustice and justice are relevant to the discourse for technology and policy, and how they can inform better practices for the designers. 

\section{Epistemic Injustices}
Philosopher Miranda Fricker defines epistemic injustice as "a wrong done to someone specifically in their capacity as a knower" \cite{fricker_epistemic_2011}. Fricker further identifies two types of epistemic injustice: testimonial injustice and hermeneutical injustice, both of which are particularly relevant for our case. Testimonial injustices occur when a person’s testimony is not trusted or given less credibility due to social biases or prejudices against the person or the social group they belong to. In other words, the person seems sufficiently credible; they suffer from a credibility deficit. Hermeneutical injustice occurs when a person does not possess the knowledge to communicate their experience due to them not possessing the shared vocabulary to express it. For instance, a non-native speaker might lack sufficiently technical or legal vocabulary in English to persuasively articulate their complaint.

Along with Fricker, others such as Jose Medina, Gaile Polhaus, Micheal Sullivan and Ian James Kidd have documented how epistemic injustices occur in various situations, experiences and areas of life such as medicine, law and education, thereby privileging the knowledges of certain groups over others \cite{michael_sullivan_epistemic_2017,carel_epistemic_2017, kidd_routledge_2017}. These injustices are a critical limitation to creating a more equitable and just society in which we can account for the experiences of those that are different from us. Within the HCI community, there is limited work on how technologies can perpetuate or even account for epistemic injustices. This paper  attempts to introduce epistemic injustice as a significant concern and an area of inquiry that is relevant for HCI, Law and Policy. 

For our case, epistemic injustice explains both, how and why the tenants’ heating complaints do not get resolved. We documented through a series of interviews with attorneys, Heat Seek and legal literature review, the process of making a heat complaint in New York. We show how the tenants’ testimonies are not believed at various points while making the heating complaint. Instead, the tenants are actively disregarded because they suffer from a credibility deficit which partly stems from their social standing. While inadequate heat is a pervasive problem across all five boroughs of the city, some neighborhoods, such as the Bronx, are disproportionately affected each year. These are the places where marginalized and minority communities live, and the complaints from these areas make up a bulk of the heating complaints \cite{renthop_bronx_2022}.

\section{Heating Complaint}

Each year, the Housing and Preservation Department (HPD) receives in excess of 150,000 heating complaints \cite{housing_preservation_and_development_new_york_city_department_of_heat_2022}. Most of these complaints are closed quickly \cite{new_york_city_comptroller_turn_2023}. For instance, we found that once HPD has received a complaint, it will follow up with the landlord and close the complaint based on the landlord’s word regardless if the heating has been restored or not. Here, the landlord’s word is assigned a higher epistemic value than the tenant’s complaint, i.e. the tenant suffers from a credibility deficit. Similarly, a review of the court documents illustrates how the court itself discounts certain instances of the tenant’s experience because they are not specific enough \cite{supreme_court_appellate_term_new_york_stellar_2019}. 

Once HPD receives the complaint, and it has not been closed based on landlord's responses, HPD will send a building inspector to assess the legitimacy of the complaint. If there is no heat, the inspector can issue a heating violation that can then be used as a basis for a court proceeding. Unfortunately, getting this violation is also a difficult task. For instance, before the building inspector visits the apartment, they will have to inform the landlord but not the tenant. Landlords tend to turn up the heating for the impending visit, only to turn it off once the inspector has left, thus avoiding the violation (and also costing the city money). Also, as the tenant is not usually informed of the visit, they have to accommodate this surprise visit in their daily routines, which is especially strenuous for people working multiple jobs, performing caretaking duties or are differently-abled. Even when the inspector is able to access the apartment, they might not issue the heating violation if the outside temperature is not low enough on that specific day, meaning that if the inspector visits on an unusually hot day or time, there will be no violation, regardless of how much time the tenant has spent without heat. Often, the tenant's testimony is systemically disregarded in the favor of the landlord's, during the inspection process, simply because they are "tenants" -- a social and economic class \cite{michener_politics_2022}. We can trace such instances of tenants being subjected to testimonial and hermeneutical injustice all the way to the workings of the Housing Courts.

Susana Blankely, a community organizer in New York, documents the tenant’s experience in a housing court. Here, the tenants are frequently cornered by the landlord’s attorney, are misinformed about their cases or even completely dismissed \cite{blankley_fight_2016}. For most of the tenants, they have to navigate this process alone without any help from an attorney because they usually are unable to afford one. Being asked to conduct a fairly complex legal process without any experience or guidance is an example of hermeneutical injustice. Proving any effect in the court of law requires "specificity of dates and the relative severity" of the inadequate heat \cite{supreme_court_appellate_term_new_york_stellar_2019}. There are specific rules of evidence, that establish the evidence to be reliable and verifiable. Thus, the inability to prove to lack of heat, an inherently embodied and physical experience, partly due to lack of documentation and partly due to limited understanding for the rules of evidence, the tenants suffer from both testimonial and hermeneutical injustices. There are other such instances where the tenant is constantly and intentionally wronged in their capacity as a knower, simply due to their identity as a tenant, and a person of color, a specific gender or even age. Policy solutions such as HPD’s protocols for handling complaints and inspections, or even the temperature limit itself, do not account for the credibility differentials between individuals and social groups. This creates a situation which fails the people that need help the most. Heat Seek's response and its success is further evidence of epistemic injustice.

\begin{figure}
    \centering
    \includegraphics[width=\linewidth]{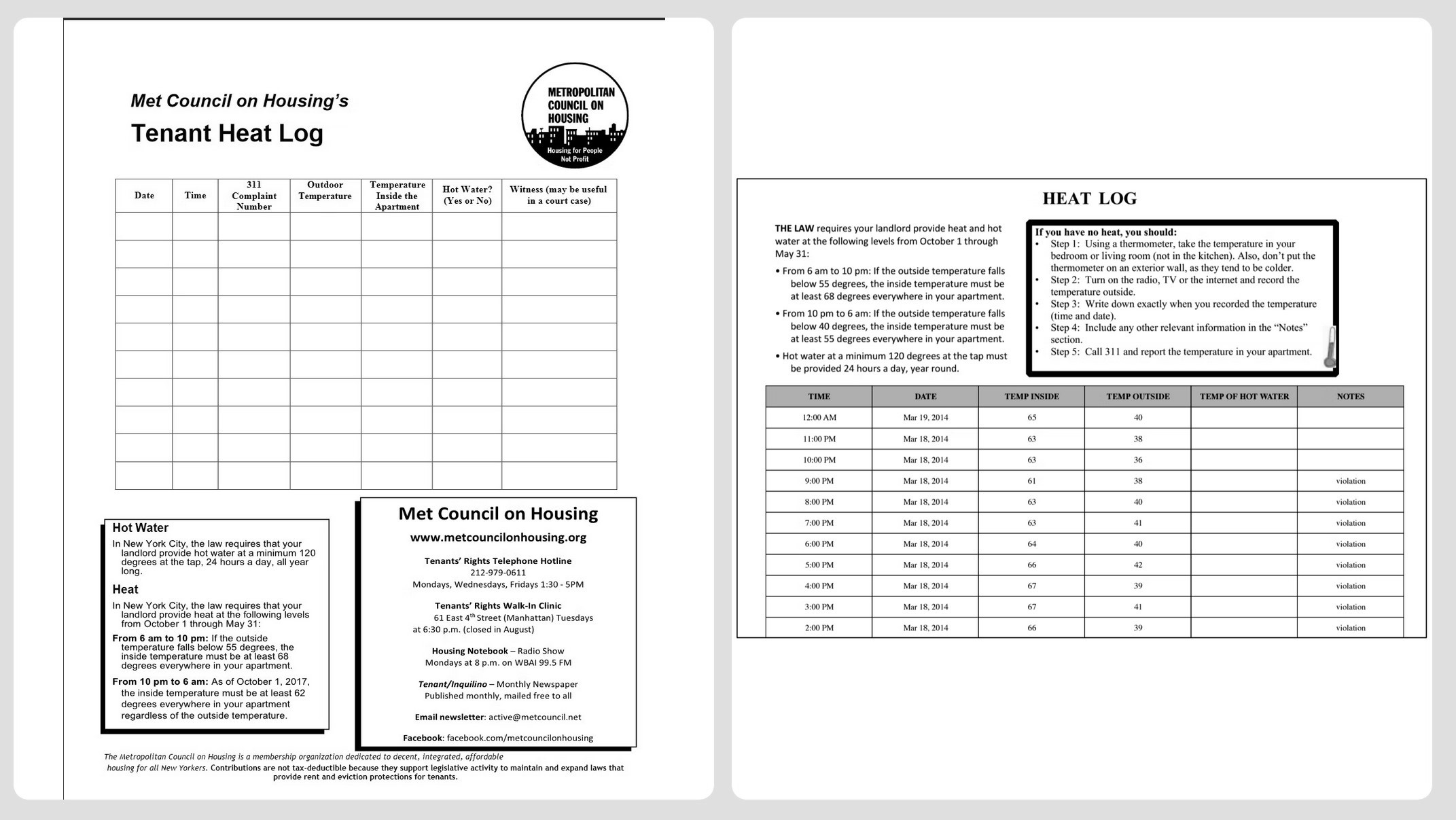}
    \caption{The manual log (left) vs the Heat Seek log (right)}
    \label{fig:log}
    \Description{A table of temperature log which include data for date, time, temperature inside, temperature outside, hot water and notes}
\end{figure}

\section{Heat Seek}
Heat Seek’s temperature sensor is a simple piece of technology. It is a temperature sensor connected to the internet that collects hourly readings of the inside temperature, pulls the outside temperature from the internet, documents the date and time, and calculates if the inside temperature is in violation or not of the legal limit. Tenants, and their attorneys, can access the logs and use it to monitor compliance with the temparture limits. This simple tool has been extremely effective in making sure landlords keep the heat on to at least the minimum limits. Based on our interviews and news reports, Heat Seek has helped tenants who had been without heat for years and other recourse had failed them. 

Examining t it through the lens epistemic injustice, we can see why Heat Seek is so effective. By providing the tenants with quantitative, tangible data in the form of temperature logs, Heat Seek has made the tenants more credible. Instead of relying solely on the testimonies of tenants against the landlord, now the court and attorneys can also assess the heating log produced by a third party. Heat Seek actively positions itself as an “objective” and “neutral” party to the situation \cite{heat_seek_heat_2021}. This lends Heat Seek -- and by proxy, tenants -- increased credibility when making heat complaints. All data, as reported by Heat Seek and corroborated by interviews with attorneys, points to the fact that tenants are taken more seriously when they rely on Heat Seek’s data. Heat Seek has also partnered with community organizations and legal aid clinics to provide the sensors to the people who can benefit the most from them.

\begin{figure}
    \centering
    \includegraphics[width=\linewidth]{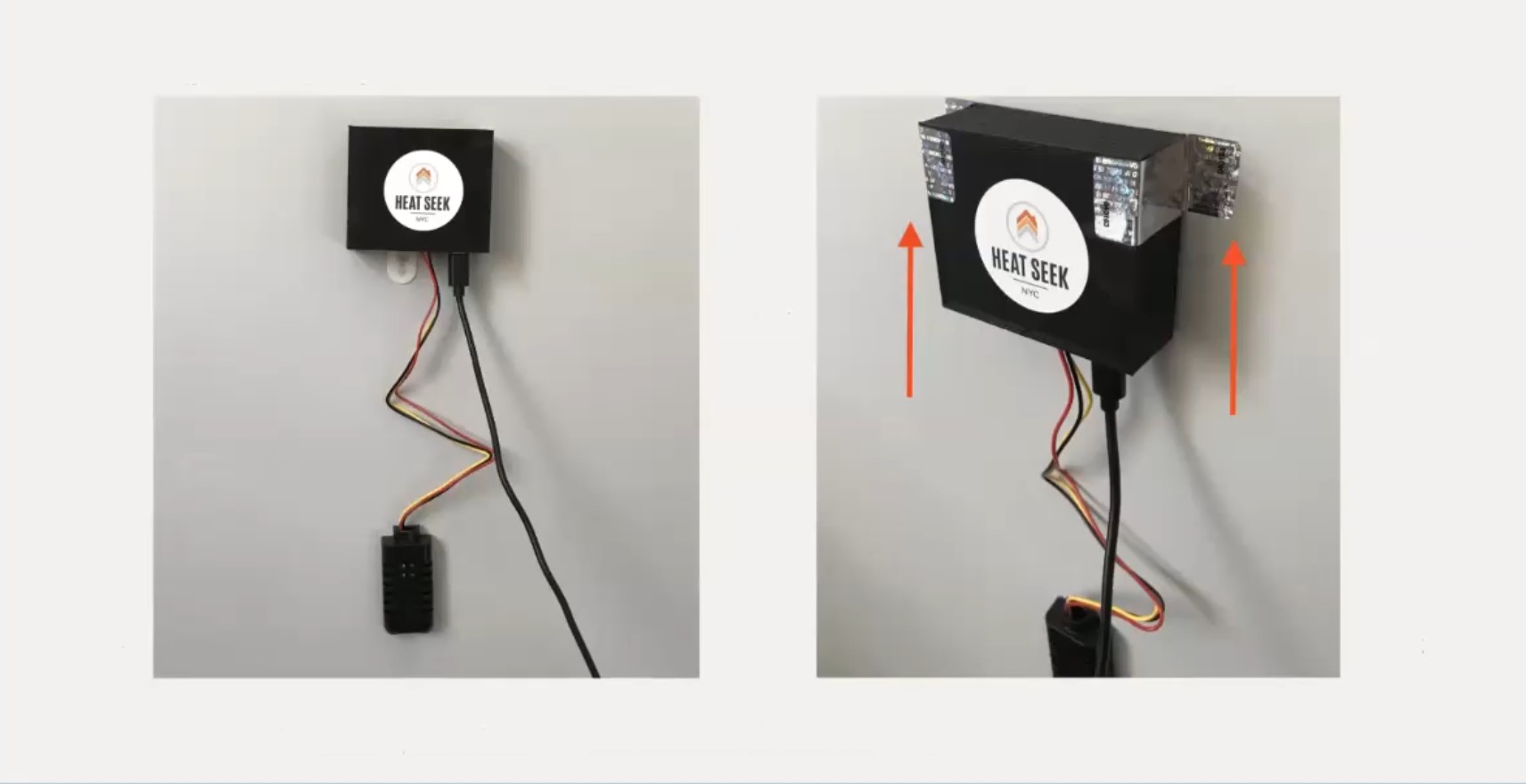}
    \caption{Heat Seek's sensor placed on a wall. The tamper-evident tape is visible in the image on the right. Image Courtesy of Heat Seek}
    \Description{A temperature sensor box stuck on a white wall and connected to a power outlet}
    \label{fig:sensor}
\end{figure}

Before Heat Seek, tenants making a heat complaint were asked by their attorneys to build a temperature log \cite{metropolitan_council_on_housing_heat_2023}. Interestingly, the epistemic move becomes clear when one considers that the data from Heat Seek is by all means identical to the one that a tenant would fill in their own log. Both these logs have the same data points (see fig 1), only differing in the frequency of data. However the manual logs were not nearly as effective as the logs from the Heat Seek sensor in getting the heat restored. The manual logs seem to possess a lower epistemic value because the tenant that produced them suffers from a credibility deficit, unlike Heat Seek which is a “legitimate”, “credible” and a neutral third party. While on surface it might appear as if the tenant’s epistemic value has increased through the credibility boost from the presence of Heat Seek, a deeper look reveals that their epistemic standing remains the same; only by relying on the credibility of a “technofix” the tenants can receive a desired outcome, and those without it carry on with the same fate. This not to criticize how Heat Seek operates, but to illustrate the severity of epistemic injustice, that even interventions that provide credibility boosts can not entirely alleviate the credibility deficits and the boosts might even be temporary and ad-hoc. 

\section{Discussion \& Conclusion}

Analyzing Heat Seek through the lens of epistemic injustice illustrates the critical and generative capacity that the examinations of epistemic injustice offers for the design and policy interventions. By examining the case of making a heat complaint in New York from the perspective of epistemic injustice, one can identify the points and interactions where the process fails, such as when the complaints are closed based on landlords' word or when the inspectors do not inform the tenants of an inspection. These failures provide an opportunity to design interventions such as Heat Seek. Understanding Heat Seek as an attempt to bring epistemic justice allows us to examine how the technology operates, its effectiveness and how to account for its shortcomings in different manner. This perspective emphasizes how the knowledge of certain groups and individuals, when not accounted for, can create harmful effects in the everyday lived experience of the people and even socially conscious interventions can have limited effect. 

It is important to understand that this paper is not arguing that epistemic injustice is the only concern that should be taken into consideration when designing societal level technologies. Instead, it is attempting to add epistemic injustice to the critical lens of a technologist or a policy maker. Even with the heat complaint, the issues do not arise solely from epistemic inequalities. Instead there are other structural and societal factors at play that contribute to this problem, such as landlords wanting to push tenants out, the limited design of Courts and housing shortage in urban New York. Interventions that take into account a more holistic view of the problem are going to fare better than those that do not. Heat Seek as a response also considers this scale of the problem, understanding that the tenant’s inability to get heat restored is also due to the lack of resources and help available to them. As a result, they actively partner with community organizations, legal aid clinics and tenant associations to get the sensors to the tenants that would benefit the most from them.

Civic technologies, and technologies in general, have potential to trigger policy changes. For instance, Heat Seek may seem like a small project (limited to 58 buildings from the hundreds of thousands in New York), but it has actually enabled the City to create a temperature sensor program \cite{neubauer_proposal_2016, new_york_city_department_of_housing_and_preservation_heat_2021}. Extending considerations of epistemic injustice to this situation, we might ask if introducing sensors to all apartments is effective or even feasible? Do we run the danger of stratifying access to justice through means of technology, so those without the sensors are \textit{always} dismissed? Instead, should we not focus on improving the infrastructures in a way that respect and respond to the epistemic positions of the individuals within them?  Epistemic positions are not just critical for technology but also for policy based solutions, and perhaps more so because policies are a hermeneutical resource. 

This paper has presented the case of heating complaints as a way to invite attention to the issues of epistemic injustices prevalent around us in many forms. It seeks to engage in conversations around how these epistemic injustices are perpetuated and can be remedied by technologies and policies. We argue that a key part of designing better technologies and policies is to consider an individual as a unique knower and accounting for the limits of their knowledge. We hope that the participants at the workshop critically examine, reflect and apply the sensibilities of epistemic injustices to their own practices. 

\bibliographystyle{ACM-Reference-Format}
\bibliography{heat_seek_draft}

\end{document}